%% file: specific_heat.tex
\documentclass[twocolumn,amssymb, aps,nobibnotes,prd,showpacs]{revtex4-1}
\usepackage[pdftex]{graphicx}
\usepackage{bm,braket}
\usepackage{amsmath, amssymb}
\usepackage{color}
\usepackage{ulem}

\begin{document}
\title{Physical properties of weak-coupling quasiperiodic superconductors}
\author{Nayuta~Takemori$^{1,2}$}
\email{takemori@okayama-u.ac.jp}
\author{Ryotaro~Arita$^{2,3}$}
\author{Shiro~Sakai$^2$}
\email{shiro.sakai@riken.jp}
\affiliation{
$^1$Research Institute for Interdisciplinary Science, Okayama University, Okayama 700-8530, Japan\\
$^2$Center for Emergent Matter Science, RIKEN, Wako, Saitama 351-0198, Japan\\
$^3$Department of Applied Physics, The University of Tokyo, Hongo, Tokyo 113-8656, Japan
}
\date{\today}%

\begin{abstract}
We numerically study the physical properties of quasiperiodic superconductors with the aim of understanding superconductivity in quasicrystals.
Considering the attractive Hubbard model on the Penrose tiling as a simple theoretical model, we calculate various basic superconducting properties and find deviations from the universal values of the Bardeen-Cooper-Schrieffer theory. In particular, we find that the jump of the specific heat at the superconducting transition is about 10-20\% smaller than that universal value, in consistency with the experimental results obtained for the superconducting Al-Mg-Zn quasicrystalline alloy. 
Furthermore, we calculate current-voltage characteristics and find that the current gradually increases with the voltage on the Penrose tiling in contrast to a rapid increase in the periodic system. These distinctions originate from the nontrivial Cooper pairing characteristic to the quasiperiodic system.
\end{abstract}

\maketitle
\maketitle

\section{\label{sec:intro}Introduction}
The superconductivity in the system without translational symmetry has been 
found in amorphous metals such as Sn$_{0.9}$Cu$_{0.1}$ ($T_{\rm c}$=6.76~K)~\cite{Comberg1974,BERGMANN1976159} and Pb$_{0.75}$Bi$_{0.25}$ ($T_{\rm c}$=6.9~K)~\cite{Hasse1968,BERGMANN1976159}.
Because the ratio of the zero-temperature gap $2\Delta$ and the critical temperature $T_c$ is around 4.5 in these superconductors, they are considered to be formed by relatively strong electron-phonon interaction. On the other hand, the existence of weak-coupling superconductivity with spatially extended Cooper pairs is a highly nontrivial issue in aperiodic systems. 
Recently, an experimental work discovered bulk superconductivity in a Bergmann-type Al-Mg-Zn quasicrystalline alloy~\cite{Kamiya2018}.
The measured properties in this alloy appear to be consistent with a weak-coupling superconductor.
This discovery necessitates a theoretical investigation because the quasicrystal breaks the fundamental prerequisite in the BCS theory, namely, the presence of the momentum space and Fermi surface. 

The effect of a quasiperiodic potential on a superconductor in a 
periodic lattice has been intensively studied in one dimension 
\cite{tezuka10,tezuka13,degottardi13,cai13,ywang16,jwang16,ghadimi17}, 
especially in relation with an ultracold-atom experiment \cite{roati08}.
On the other hand, a large part has remained unexplored in two or three 
dimensions, as well as in quasicrystalline systems where the lattice points are arranged in a quasiperiodic manner; exceptions include 
Refs.~\cite{fulga16,Sakai2017,araujo19,sakai_2019,nagai2020,Cao20}.

In our previous work~\cite{Sakai2017}, we studied possible superconductivity emerging in a  
quasiperiodic system by introducing a simplified theoretical model, i.e., 
the attractive Hubbard model on the Penrose tiling~\cite{Sakai2017}. We studied the model by means of real-space dynamical mean-field theory~\cite{cold1,cold3,surface,Okamoto04,Takemori} 
and revealed that the emerging superconducting phase is categorized into three different regions, which cross over each other, in the density-interaction phase diagram. Especially, unconventional spatially extended Cooper pairs whose coherence length is much longer than the minimum length scale of 
inhomogeneity, i.e., lattice spacing, were found in the weak-coupling region, which may be relevant to the bulk superconductivity observed in the Al-Mg-Zn quasicrystal. 
We clarified that the Cooper pairs in this region deviate from that of the BCS superconductivity formed between the electrons at momentum $\bm{k}$ with spin $\uparrow$ and $\bm{-k}$ with $\downarrow$~\cite{Sakai2017}, because of the lack of the periodicity.
Moreover, the obtained superconducting states show a spatial inhomogeneity owing to the aperiodic feature of the quasiperiodic structure. Then, the self-similarity of such structures means that this superconducting state is inhomogeneous in {\it any} length scale, in distinction from any other known superconductors.

Here, a question arises: 
Does this quasiperiodic superconductor show any properties, in particular experimentally observable ones, different from those of the BCS superconductor? 
To answer this question, we calculate experimentally observable quantities such as the specific heat and current-voltage characteristics, as well as several basic quantities of the superconductors, in the Penrose tiling. We find that the jump of the specific heat is about 10-20\% smaller than that obtained by the BCS theory. Also, we find that the current ($I$) - voltage ($V$) curve shows a gradual increase
in the Penrose tiling, which is clearly different from a rapid increase in 
 the BCS theory. These results call for a further experimental investigation of these quantities in quasicrystalline superconductors.

This paper is organized as follows. 
In Sec.~\ref{sec:mandm}, we introduce the model and summarize our theoretical approach.
In Sec.~\ref{sec:results}, we show calculated results for the experimentally observable quantities such as thermodynamic properties and transport properties.
A summary is given in the last section.

\section{Model and Method}
\label{sec:mandm}

Although the present study is motivated by the recent discovery of a real quasicrystalline superconductor~\cite{Kamiya2018}, our study aims to clarify general features of quasicrystalline superconductors, rather than focusing on a specific material. To this end, we consider a theoretically tractable model containing the two essences, quasiperiodicity and superconductivity, as is done in several recent works \cite{Sakai2017,araujo19, sakai_2019,nagai2020}:
We study weak-coupling superconductivity in quasiperiodic systems by assuming a weak local attractive interaction $U<0$ in the Hubbard model on the Penrose tiling
under the open boundary condition~\cite{Takemori,Sakai2017,sakai_2019}.
Here we adopt a vertex model where a site is placed on each vertex of the rhombuses, and thus the model is bipartite. The coordination number in the system ranges from 2 to 7 and each vertex pattern can be divided into eight classes~\cite{Bruijn1,Bruijn2,Takemura,Takemori_dual}. 
The structure of the Penrose tiling is generated by applying the inflation-deflation rule~\cite{Levine1984} iteratively
to the pentagon structure composed of the five fat rhombuses so that the structure holds
 $C_{5v}$ symmetry (5-fold rotational and mirror symmetries). In the following, we make use of a series of the Penrose tiling consisting of $N=$1591, 4181 and 11006 sites. These structures contain 175, 444 and 1142 geometrically inequivalent sites, respectively. In this paper, we suppose a finite electron-transfer integral $t$ only between the vertices connected by edges of the rhombuses and set it as the unit of energy. In order to avoid a peculiarity at the half-filling, we tune the chemical potential to obtain a filling away from half-filling. In the noninteracting limit of this model, the width of the site-averaged local density of states is about 8.5$t$, so that, based on previous studies~\cite{Micnas1990,Sakai2017}, we can expect a weak-coupling superconductivity for $|U|\lesssim 4$. Because the Cooper pairs are less extended for a larger $|U|$, in order to reduce a finite-size effect coming from the boundary of the cluster, we mainly study  
$U=-3$ at quarter-filling in this paper.

To study the superconducting solution in this model, we employ the Bogoliubov-de Gennes (BdG) equation, which is expected to work well in the weak-coupling region. 
We self-consistently solve the BdG Hamiltonian given by~\cite{de2018superconductivity},
\begin{eqnarray}
\label{eq:bdg}
[{\hat H}_{\rm BdG}]_{ij}= \left[U \langle c_{i \uparrow} c_{i \downarrow} \rangle \sigma_1+\left( \frac{Un_i}{2}-\mu \right) \sigma_3 \right]\delta_{ij}\nonumber \\ 
-t\sigma_3\delta_{\langle ij \rangle}.
\end{eqnarray}
Here, $\langle ij \rangle$ denotes a pair of the neighboring sites, $\mu$ is the chemical potential and $\sigma_{1(3)}$ is $x$ ($z$)-component of the Pauli matrix. We define the site-dependent superconducting order parameter and electron density by ${\rm OP}_i=\langle c_{i \uparrow}c_{i \downarrow} \rangle$ and $n_i=\sum_\sigma  n_{i\sigma}$ with $n_{i\sigma}=\langle c_{i\sigma}^{\dagger}c_{i \sigma} \rangle$, respectively, where $c^{(\dagger)}_{i\sigma}$ is 
an annihilation (creation) operator of an electron at the $i$th site with spin $\sigma=\uparrow,\downarrow$. 
The eigenvalue $E_\alpha$ of the Hamiltonian and the local density of states enable us to compute various experimental observables. We assume only $s$-wave superconductivity driven by the local attractive interaction.

\section{Results}
\label{sec:results}

\subsection{Superconducting order parameter and gap}
\label{sec:op}
\begin{figure}[htb]
\includegraphics[width=\linewidth] {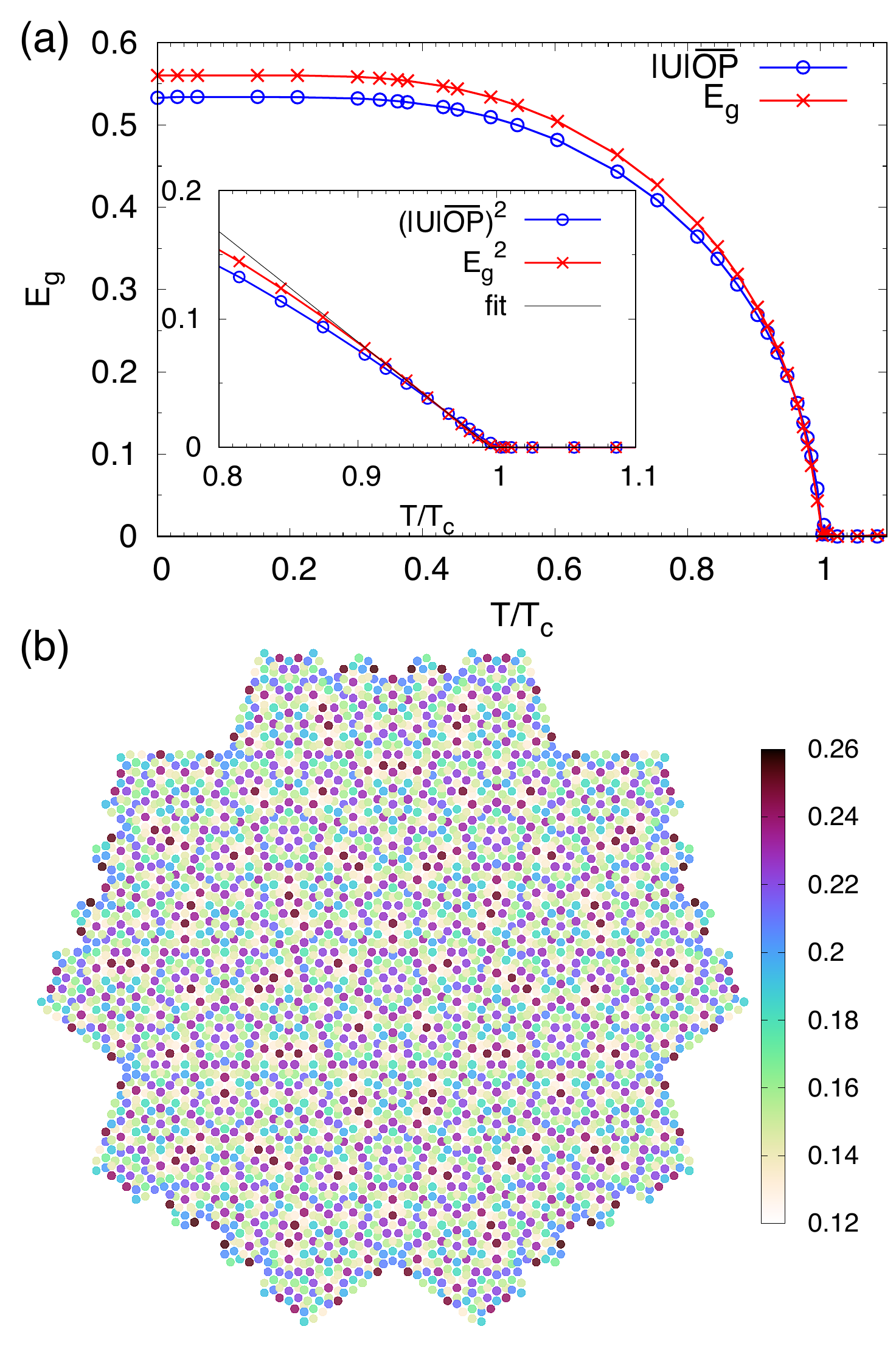}
\caption{\label{fig:OP} 
(Color Online) (a) Spatially averaged superconducting order parameter $\overline{\rm OP}$ and superconducting gap $E_g$ at quarter-filling for $U=-3$ as a function of the temperature. The inset shows square of $\overline{\rm OP}$ and $E_g$ around $T_c$ with a linear fitting function (black line). (c) Spatial pattern of site-dependent superconducting order parameter OP$_i$ at zero temperature in the Penrose tiling of 4181 sites.}
\end{figure}

\begin{table*}[htb]
  \begin{tabular}{|c|c|c|c|c|c|c|c|c|} \hline
  \multicolumn{1}{|c|}{} & \multicolumn{4}{|c|}{Penrose} &\multicolumn{2}{|c|}{Square}&  \multicolumn{1}{|c|}{BCS} \\ \hline 
     &1591& 4181& 11006 &  ext   &2500&10000 & \\ \hline 
     {\large $\frac{2E_g^0}{T_c}$ } & 3.35  & 3.38 & 3.38& 3.38    & 3.46 & 3.45&3.52  \rule[0mm]{0mm}{5mm}\\
       $A_1$ &  1.61  & 1.63& 1.69&1.70 &1.70 & 1.70&1.74 \rule[0mm]{0mm}{5mm}\\  
    {\large $\frac{\Delta C}{C_{\rm en}}$} &  1.13  & 1.21&1.21 & 1.21 	&1.40 &1.39 &1.43 \rule[0mm]{0mm}{5mm}\\
  \hline
  \end{tabular}
  \caption{\label{table:bcs} The ratio of the superconducting gap at zero temperature to the critical temperature (top), a coefficient in the temperature dependence of the superconducting gap near the critical temperature (middle), and the jump of the specific heat (bottom) obtained in Penrose tiling of 1591, 4181 and 11006 sites, as well as its extrapolated value and those for square lattice of 2500 and 10000 sites. Universal values in the BCS theory~\cite{tinkham1996introduction}
 are added in the rightmost column.}
\end{table*}

\begin{table*}[htb]
  \begin{tabular}{|c|c|c|c|c|c|c|} \hline
  \multicolumn{1}{|c|}{} & \multicolumn{1}{|c|}{$U=-2$} &\multicolumn{3}{|c|}{$U=-3$}& \multicolumn{1}{|c|}{$U=-4$}& \multicolumn{1}{|c|}{BCS} \\ \hline
     &$n=0.5$& $n=0.3$& $n=0.5$ & $n=0.7$   & $n=0.5$  & \\ \hline 
    {\large $\frac{2E_g^0}{T_c}$ } & 3.36  & 3.24 & 3.38& 3.29   & 3.42& 3.52   \rule[0mm]{0mm}{5mm}\\
       $A_1$ & 1.56  &1.69 & 1.69&1.51 &1.67& 1.74 \rule[0mm]{0mm}{5mm}\\  
    {\large $\frac{\Delta C}{C_{\rm en}}$} &  1.23  & 1.25&1.21 &   0.72 & 1.23 &1.43 \rule[0mm]{0mm}{5mm}\\
  \hline
  \end{tabular}
  \caption{\label{table:un} The same quantities as in Table.~\ref{table:bcs} obtained in the Penrose tiling of 11006 sites for different values of $U$ and the average filling $n$. Universal values in the BCS theory~\cite{tinkham1996introduction}
 are added in the rightmost column.}
\end{table*}

We first discuss the temperature dependence of the superconducting order parameter and the gap, using a Penrose-tiling cluster of 4181 sites.
We show in Fig.~\ref{fig:OP}(a) spatially averaged order parameter $\overline{\rm OP}\equiv \frac{1}{N}\sum_{i}{\rm OP}_i$ and superconducting gap $E_g$ which is defined as the minimum absolute value of the eigenvalues \{$E_\alpha$\}. 
As discussed in our previous paper~\cite{Sakai2017}, the local superconducting order parameter OP$_i$, which shows a non-uniform spatial distribution [Fig.~\ref{fig:OP}(b)], reaches zero simultaneously everywhere at the critical temperature $T_c$ where both $E_g$ and $\overline{\rm OP}$ 
vanish. 
In the periodic system, according to the BCS theory, the critical behavior of the superconducting gap satisfies 
\begin{equation}
\label{eq:gap}
\frac{E_g(T)}{E_g^0}\sim A_1 \left(1-\frac{T}{T_c}\right)^{\gamma},
\end{equation}
with the exponent $\gamma=1/2$. Here, $E_g^0$ denotes the superconducting gap at zero temperature. To see the critical behavior in the quasiperiodic system where the distribution of the superconducting gap is inhomogeneous, we plot the  
temperature dependence of ${E_g}^2$ in the inset of Fig.~\ref{fig:OP}(a). It shows a 
 linear behavior around $T_c$, which indicates that Eq.~(\ref{eq:gap}) is satisfied also in the present mean-field-type calculation for the quasiperiodic (inhomogeneous) system. Furthermore,
in the framework of the BdG theory, the superconducting gap is interpreted as $E_g\sim|U|\overline{\rm OP}$. Indeed, these two quantities are in good agreement around $T_c$ in the present case, too. 
By using this relationship, the critical temperature is evaluated 
as $T_c=0.330$ for the parameters used here.

Next, we compare the calculated results with the known value of the superconducting gap in the BCS theory as shown in Table~\ref{table:bcs}. 
The ratio  $2E_g^0/T_c$ is 3.52 in the BCS theory. This is nearly reproduced 
with the present method applied to a square lattice of finite sizes 
although the calculated value is slightly smaller than the BCS one.
On the other hand, the ratio is calculated to be 3.38 for 
the Penrose tiling.
Here, we have calculated the ratio for each finite-size cluster and the extrapolation to the thermodynamic limit in the Penrose tiling gives 3.38, which is substantially smaller than 
that of the BCS theory.
This substantial reduction from the BCS value 
can be a notable characteristic of the quasiperiodic superconductivity since the effect of a finite coupling (as we use $U=-3$) usually lifts the ratio from the BCS value~\cite{Bak98,Toschi05}, in contrast to the present case. This indicates that distinct weak-coupling superconductivity is formed in the present system.

For the temperature dependence of the superconducting gap near the critical temperature, 
the coefficient $A_1$ in Eq.~(\ref{eq:gap}) in the BCS theory is given by $A_1=1.74$. We obtain a similar value both for the Penrose tiling and square lattice, as shown in Table~\ref{table:bcs}. These results show that the critical behavior, which is scaled by $E_g^0$, does not show much difference between the Penrose tiling and the square lattice. 

To examine whether these results are universal in the Penrose tiling, we have performed calculations at different average fillings $n$ and interactions for $N=11006$ as summarised in Table.~\ref{table:un}. We see that $2E_g^0/T_c$ is always smaller than the BCS value. $A_1$ is similar to or somewhat smaller than the BCS one. Note that the finite-size effect may influence the results at $U=-2$ while a strong-coupling feature (related to the Bose-Einstein condensation) \cite{Micnas1990} may come in the results at $U=-4$.

Thus, 
we clarified that $2E_g^0/T_c$ exhibits a smaller value in the Penrose tiling than in the periodic system, while the temperature dependence of the superconducting gap does not show a clear difference from that in the BCS theory.

\subsection{Perpendicular space}
\begin{figure*}[htb]
\includegraphics[width=\linewidth] {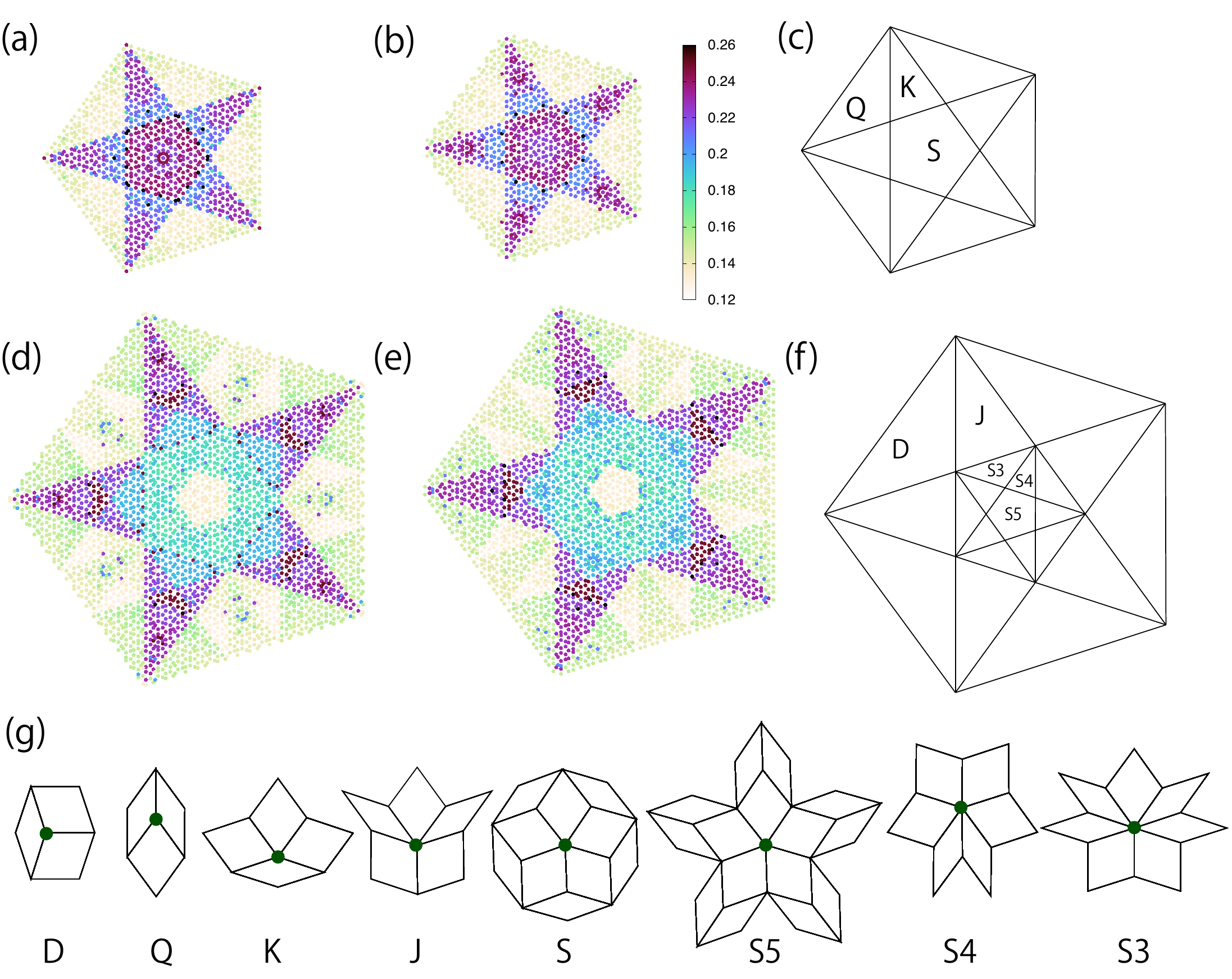}
\caption{\label{fig:perp_OP} 
(Color Online) Perpendicular space profile of the site-dependent superconducting order parameter OP$_i$ for (a)~$z_{\rm{perp}}=0$, (d)~$z_{\rm{perp}}=1$, (e)~$z_{\rm{perp}}=2$ and (b)~$z_{\rm{perp}}=3$
at zero temperature in the Penrose tiling of 11006 sites. Each perpendicular space for 
(c)~$z_{\rm{perp}}=0, 3$ and (f)~$z_{\rm{perp}}=1, 2$ 
is divided into star-shaped sections which correspond to the vertex pattern. (g) Vertex patterns D, Q, K, J, S, S5, S4 and S3 defined for the Penrose tiling~\cite{Bruijn1,Bruijn2}.}
\end{figure*}
The superconductivity in the weak-coupling 
region involves spatially extended Cooper pairs~\cite{Sakai2017}, where the off-site superconducting order parameter $\langle c_{i\uparrow}c_{j\downarrow}\rangle$ remains finite for a large distance between $i$ and $j$. In this state, the local superconducting order parameter reflects the geometry beyond the nearest neighbors. This feature can be easily seen in perpendicular space~\cite{KogaTsunetsugu} as shown in Fig.~{\ref{fig:perp_OP}}. Perpendicular space ($x_{\rm perp}$, $y_{\rm perp}$, $z_{\rm perp}$) is the remaining three dimensions when the projection from a five-dimensional cubic lattice onto two-dimensional physical space generates the Penrose tiling~\cite{senechal}. In this space, a parity of $z_{\rm perp}\in \{0,1,2,3\}$ corresponds to the sublattice in physical space. Moreover, 
sites with equivalent local vertex geometries~\cite{Takemura,Bruijn1,Bruijn2} in physical space are assembled in the star-shaped section in the perpendicular space as shown in Figs.~\ref{fig:perp_OP}(c) and (f). 
Therefore, the roughly uniform color in each section indicates that the value of ${\rm OP}_i$ is largely determined by the local geometries represented by the vertex patterns in Fig.~\ref{fig:perp_OP}(g). However, a closer look at 
Figs.~\ref{fig:perp_OP}(a), (b), (d) and (e) does not show merely a star-shaped pattern but further additional structures, which indicate that longer-range geometry beyond the nearest neighbors plays a role. We note that the points with exceptionally strong intensity in the D region [Figs.~\ref{fig:perp_OP}(d) and (e)] correspond to the sites at the edge of the system.

\subsection{Local density of states} 
\begin{figure}[htb]
\includegraphics[width=\linewidth]{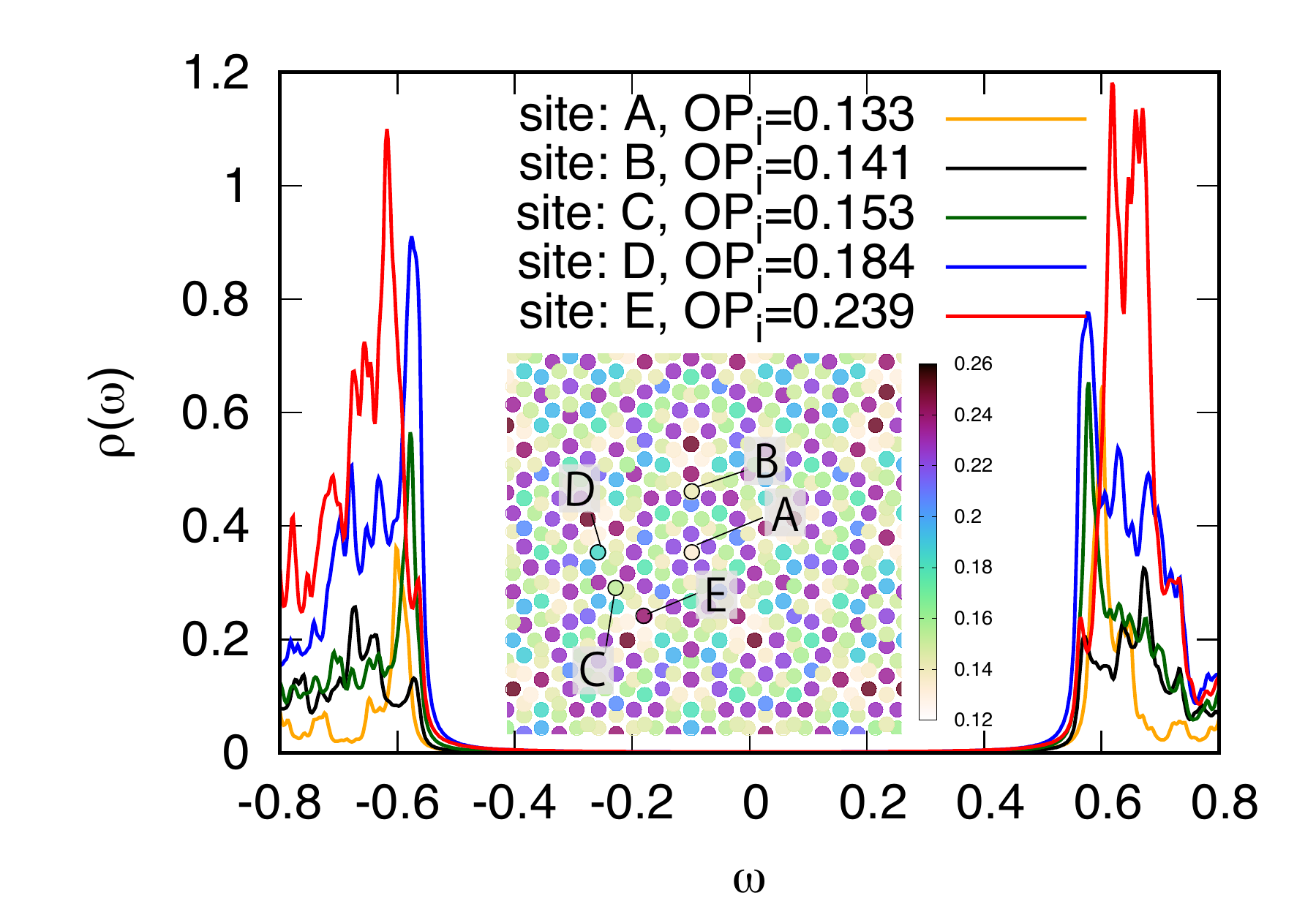}
\caption{\label{fig:STS} 
(Color Online) Local density of states for geometrically inequivalent sites at zero temperature obtained for the Penrose tiling of 4181 sites at quarter-filling for $U=-3$. The inset shows enlarged view of Fig.~\ref{fig:OP}(b) around the center of the Penrose-tiling cluster. To plot the density of states, we have added an imaginary part $i\eta$ (with $\eta=0.003$) to the energy $\omega$.}
\end{figure}

As shown in Fig.~\ref{fig:OP}(b), the superconducting order parameter shows an interesting nonuniform spatial pattern in quasiperiodic systems. This may be observable by scanning tunneling microscopy (STM) or scanning tunneling spectroscopy (STS) as a gap in the local density of states. To clarify this point, we calculate the local density of states as shown in Fig.~\ref{fig:STS} for five different sites A-E depicted in its inset. 
These sites are geometrically inequivalent and show different values of local superconducting order parameter, ${\rm OP}_i$=0.133 (A), 0.141 (B), 0.153 (C), 0.184 (D), 0.239 (E).
Although the local superconducting order parameter depends significantly on sites, 
the gap size 
in the local density of states does 
not appreciably depend on ${\rm OP}_i$. 
On the other hand, the amplitude of the peaks (Bogoliubov peaks) at the edge of the gap strongly depends on sites and this would be measurable by STM/STS. Although the site dependence of this peak amplitude does not seem to show a simple correspondence to that of ${\rm OP}_i$, it 
can still provide evidence of the inhomogeneous superconducting state characteristic to quasiperiodic systems and would show a fractal-like pattern in real space.
In the next section, we shall discuss how this spatial inhomogeneity affects the bulk superconducting property. 

\subsection{Specific heat}
\label{sec:specific}
 
\begin{figure}[htb]
\includegraphics[width=\linewidth]{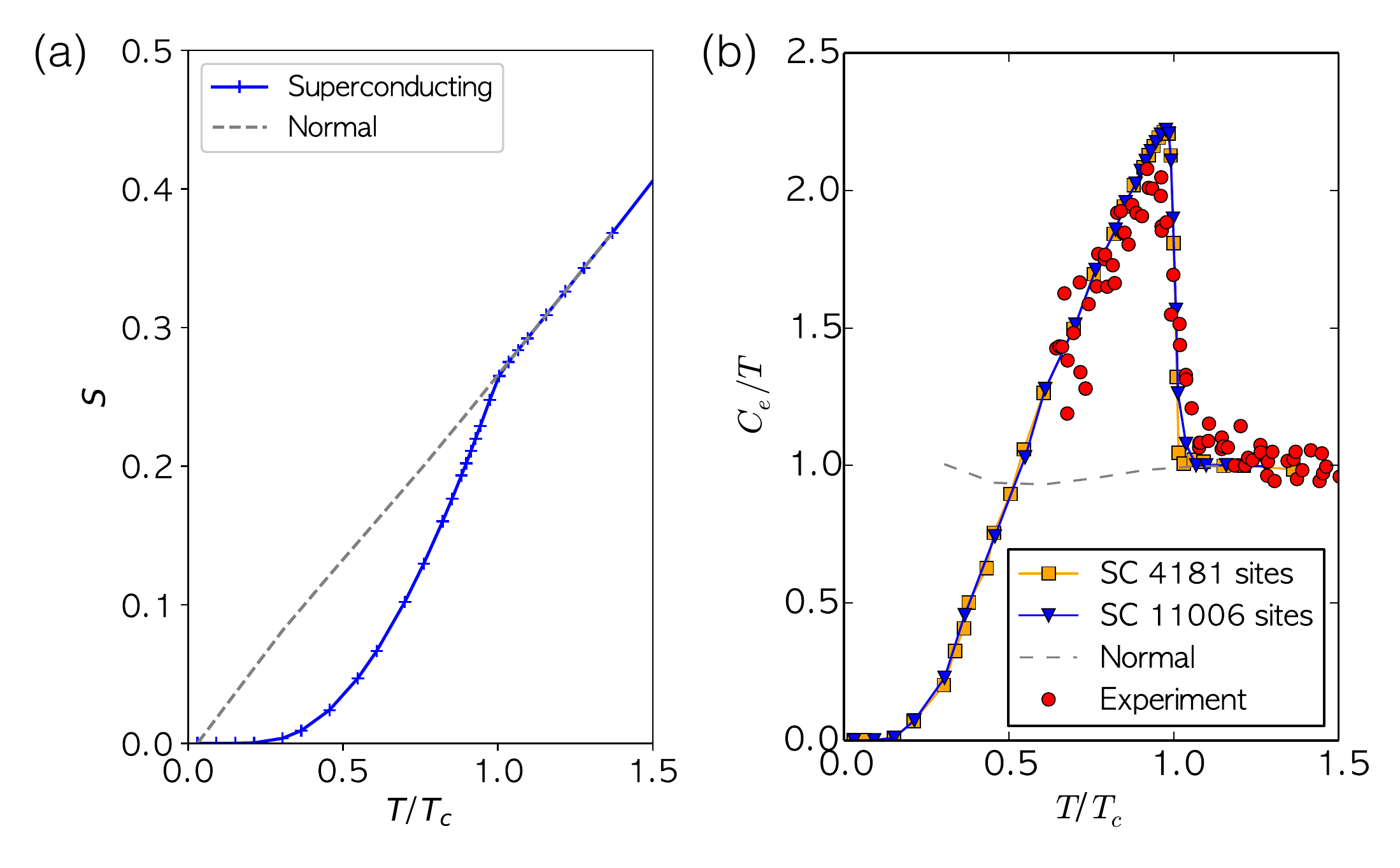}
\caption{\label{fig:e-s} (Color Online) Temperature dependence of entropy $S$ (a) and specific heat $C_{e}/T$ (b) obtained in the Penrose tiling of 11006 sites at quarter-filling for $U=-3$. We note that the specific heat is shown in the unit of $C_{en}/T_c$ where $C_{en}$ denotes the specific heat in the normal state at $T_c$. The specific heat obtained for 
4181 sites and in the experiment for Al-Mg-Zn quasicrystalline alloy~\cite{Kamiya2018} are plotted for comparison. Dashed curve represents the results calculated for 11006 sites with artificially restricting a solution to the normal state.
}
\end{figure}

Another basic property of a superconductor is specific heat.
To obtain this quantity, we first calculate the entropy in the Penrose tiling by
\begin{eqnarray}
\label{eq:entropy}
S=2\sum_{\alpha}\left\{ \ln (1+e^{-\beta E_{\alpha}})+\frac{\beta E_{\alpha}}{e^{\beta E_{\alpha}+1}} \right\},
\end{eqnarray}
where $\alpha$ runs over 1 to $2N$ and $\beta$ denotes inverse temperature.
As the finite superconducting gap $E_g$ appears at $T<T_c$, entropy shows a kink at $T=T_c$ as shown in Fig.~\ref{fig:e-s}(a).
The electronic specific heat is obtained by numerical differentiation of entropy as
\begin{equation}
C_e=T\frac{dS}{dT},
\end{equation}
which shows a jump singularity at $T=T_c$ as shown in Fig.~\ref{fig:e-s}(b), corresponding to the kink appearing in entropy. 
In the BCS framework, the universal ratio $\Delta C/C_{\rm en}=1.43$ is known between the specific heat jump $\Delta C$ and its value in the normal state $C_{\rm en}$ at $T=T_c$.

The system size dependence of the heat capacity jump is shown in Table~\ref{table:bcs}. In the Penrose tiling, it depends on the system size only weakly and the extrapolated value is 1.21. Table~\ref{table:un} shows a similar reduction from the BCS value for other fillings and interactions, too. On the other hand, that in the square lattice shows a value close to the known BCS value 1.43.
We thus find that the jump of the specific heat in the Penrose tiling is about 10-20\% smaller than that obtained by the BCS theory.
This reduction is consistent with the experimental results of Al-Mg-Zn quasicrystalline alloy~\cite{Kamiya2018}.
The reduction of the jump in the quasiperiodic system is presumably due to the multigap nature (inhomogeneity) of its superconductivity (as is seen in Fig.~\ref{fig:STS}), which would broaden the singularity.

\begin{figure}[htb]
\includegraphics[width=0.5\linewidth] {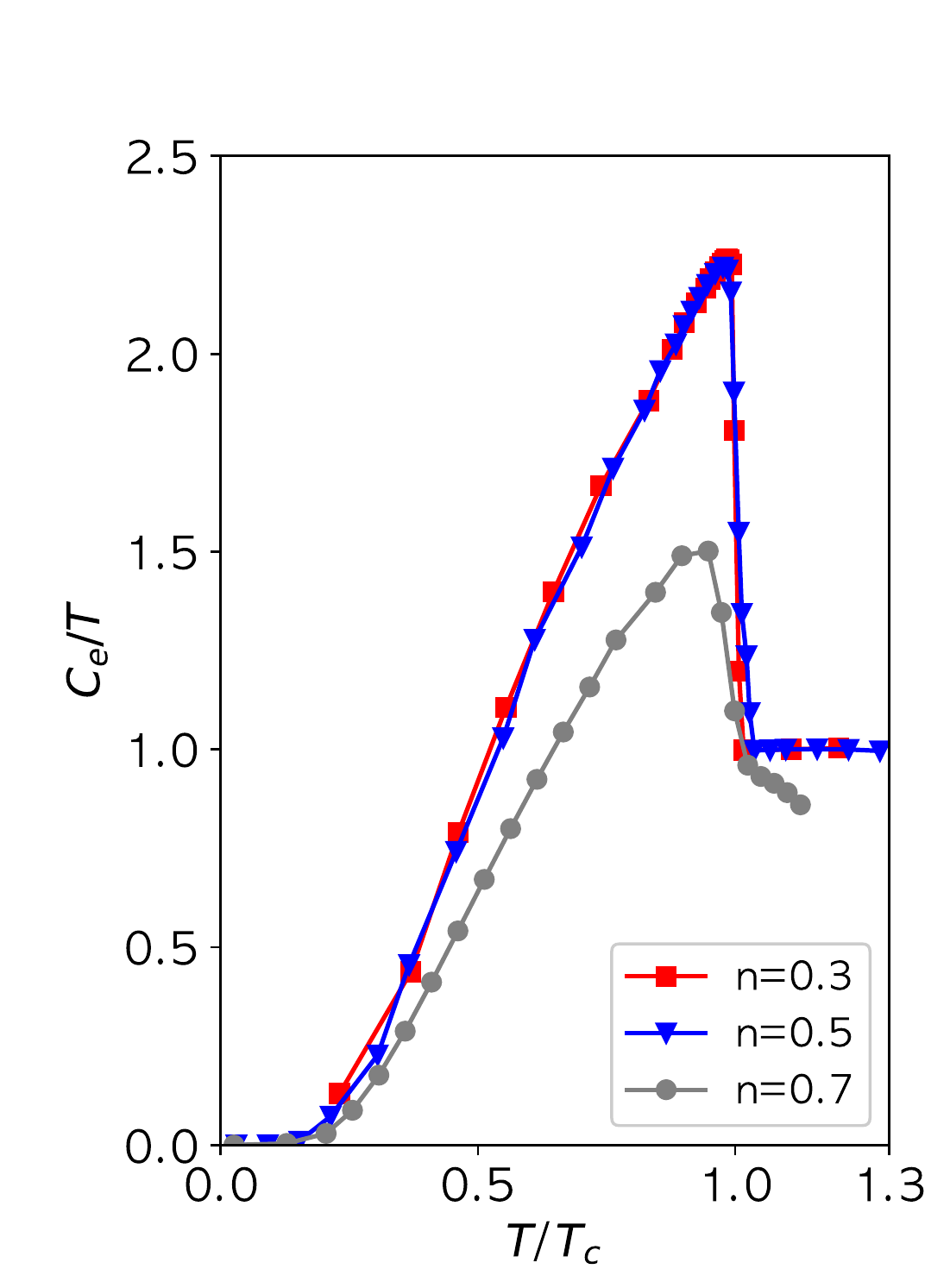}
\caption{\label{fig:sparam} 
 (Color Online) Temperature dependence of specific heat $C_{e}/T$ obtained in the Penrose tiling of 11006 sites at different values of the average filling $n$ for $U=-3$. We note that the specific heat is shown in the unit of $C_{en}/T_c$ where $C_{en}$ denotes the specific heat in the normal state at $T_c$.}
\end{figure}

A closer look at Table \ref{table:un} tells us that the jump decreases as $n$ approaches the half-filling at $U=-3$. This is consistent with the $U$-$n$ phase diagram in Ref.~\cite{Sakai2017}, which shows that the spatial extension of the Cooper pairs is suppressed around the half-filling in the weak-coupling region. Namely, a strong-coupling effect (leading to the Bose-Einstein condensation) would account for this trend. Note that the specific heat in the normal state ($T>T_c$) strongly depends on temperature for $n=0.7$, differently from the temperature-independent behavior for $n=0.3$ and $0.5$, as shown in Fig.~\ref{fig:sparam}.


\subsection{$I-V$ characteristics}
\label{sec:i-v}

\begin{figure}[htb]
\includegraphics[width=\linewidth]{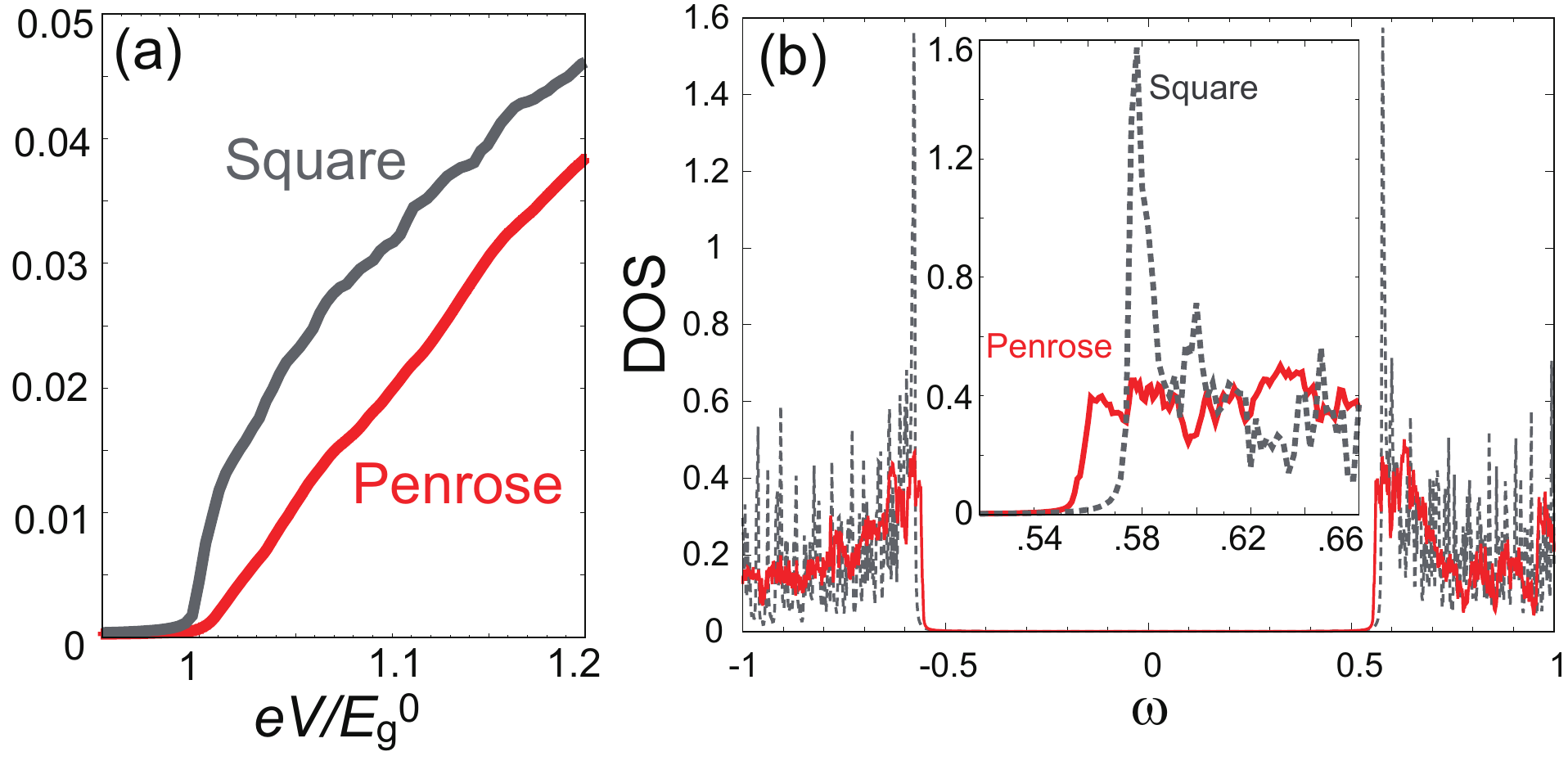}
\caption{\label{fig:i-v} 
(Color Online) (a) Current as a function of $eV/E_g^0$ and (b) the site-averaged density of states at zero temperature obtained for the Penrose tiling of 11006 sites and for a square lattice of 10000 sites at quarter-filling for $U=-3$ and $\eta=0.001$. Inset shows the enlarged view around $\omega=E_g^0$.}
\end{figure}
\begin{figure*}[htb]
\includegraphics[width=\linewidth]{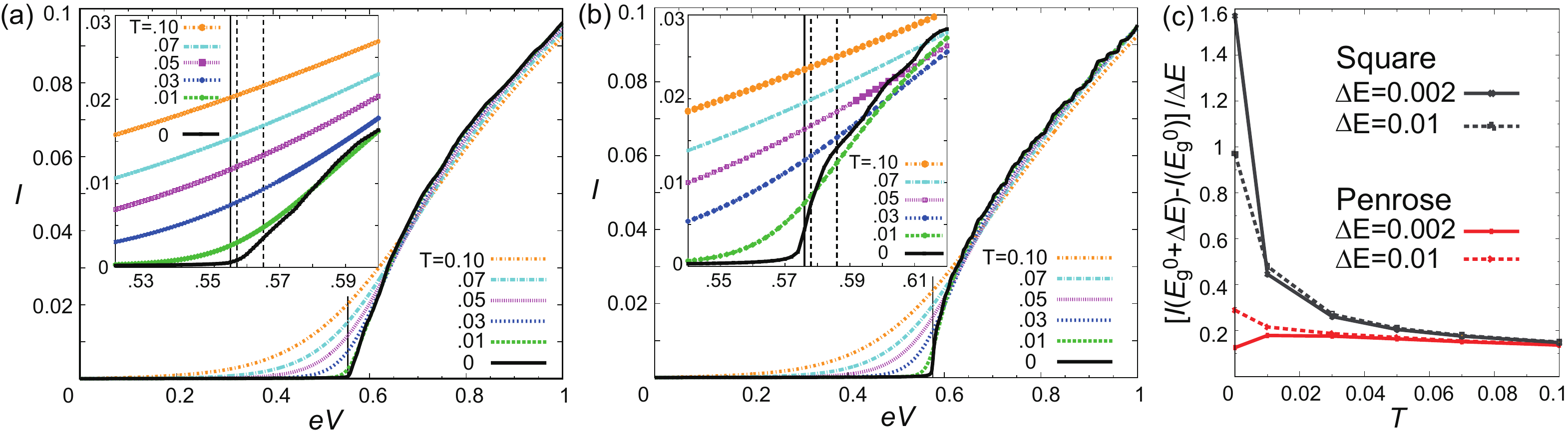}
\caption{\label{fig:i-v2} 
(Color Online) $I-V$ characteristics at various temperatures for the (a) Penrose ($N=11006$) and (b) square ($N=10000$) lattices, calculated for the quarter-filling and $U=-3$. Insets show the enlarged views around the threshold voltages, $eV=E_g^0=0.555$ and $0.576$ (denoted by solid vertical lines), respectively. The dashed vertical lines denote the voltages $eV=E_g^0+\Delta E$ with $\Delta E=0.002$ and $0.01$, which are used to calculate the slope. (c) The temperature dependence of the slope at $eV=E_g^0$ for the Penrose and square lattices. }  
\end{figure*}

Next, we focus on transport property, which might be a direct clue to distinguish the BCS superconductivity and quasiperiodic superconductivity. For this purpose, we calculate $I-V$ characteristics curve of a normal metal-superconductor tunnel junction as shown in Fig.~\ref{fig:i-v}(a). Here, we consider a tunnel junction of a periodic normal metal and a quasiperiodic superconductor. The current through the junction $I(V)$ at temperature $T=1/\beta$ is given by~\cite{tinkham1996introduction}:
\begin{eqnarray}
\label{eq:i-v}
I(V) \propto \int_{- \infty}^{\infty}\rho(E)[f(E)-f(E+eV)]dE
\end{eqnarray}
where $f(E)=1/(e^{\beta E}+1)$ is the Fermi-Dirac distribution function and $\rho(E)$ denotes the site-averaged density of states of the superconductor. We have assumed that the density of states of the normal metal does not depend on energy in the range of our interest. 

At $T=0$, the current flows only when the applied voltage exceeds the difference of chemical potentials between the normal metal and the superconductor, namely $I(V)$ becomes finite only for $e|V|\geq E_g^0$. 
In the periodic system, the voltage dependence of the tunneling current shows a rapid 
rise because of the sharp Bogoliubov peak at the edge of the superconducting gap in the density of states. On the other hand, it shows a gradual (nearly linear) development in the quasiperiodic system owing to the non-uniform distribution of the superconducting order parameter, which is reflected in the multiple peaks and a nearly flat distribution for $\omega\gtrsim E_g^0$, in the site-averaged density of states shown in Fig.~\ref{fig:i-v}(b). 

Figures \ref{fig:i-v2}(a) and (b) show the $I-V$ characteristics at various temperatures for the Penrose and square lattices, respectively. Both curves look qualitatively similar at finite temperatures. However, we find that the two cases can be distinguished by looking at the temperature dependence of the slope at the threshold voltage, $eV=E_g^0$. Here, we define the slope by the difference of $I$ at $eV=E_g^0$ and at $E_g^0+\Delta E$. As the temperature decreases, the slope increases rapidly on the square lattice while it increases only weakly on the Penrose lattice. This is plotted for $\Delta E=0.002$ and $0.01$ [denoted by vertical lines in the insets to Fig.~\ref{fig:i-v2}(a) and (b)] in Fig.~\ref{fig:i-v2}(c). The weak increase ({\it i.e.,} nearly flat behavior) in the quasiperiodic system is due to the nearly flat distribution of the Bogoliubov peaks as discussed above. For $\Delta E=0.002$, the slope even decreases from $T=0.01$ to $T=0$. This will be because the density of states lacks the sharp Bogoliubov peak at the gap edge [inset to Fig.~\ref{fig:i-v}(b)]: Since at $T=0$ the current $I$ is contributed by the spectra within the width $\eta$ ($=0.001$ here) around the gap edge, the small spectral weight at the gap edge gives a relatively small increase of $I$ compared to that at finite $T(>\eta)$ where the spectra within the width $\sim T$ around the gap edge can contribute.

These results clearly show the difference between the BCS superconductivity and quasiperiodic superconductivity, offering possible 
experimental tests to examine whether the superconductivity found in quasicrystals is consistent with quasiperiodic superconductivity or not.

\section{Conclusions}

 We calculate the experimental observables such as specific heat and current-voltage characteristics in the Penrose tiling and compare them with the well-known results in the BCS theory. We find that the specific heat jump is about 10-20\% smaller than that obtained by the BCS theory, in consistency with the experimental results obtained in the superconducting Al-Mg-Zn quasicrystalline alloy.
We also find that the ratio of the zero-temperature gap and $T_c$ is {\it smaller} than the BCS value. This is in sharp contrast with the amorphous superconductors, which usually shows the ratio substantially larger than the BCS value, indicating that the quasiperiodic superconductivity is indeed formed in the weak-coupling mechanism rather than the strong-coupling one.
These tendencies do not depend on the electron density in the weak-coupling region.
 Furthermore, we calculate current-voltage characteristics and find that the gradual, nearly linear, development appears in the Penrose tiling in comparison to that in the periodic system. These properties mark the quasiperiodic superconductivity in distinction from the BCS superconductivity.

\begin{acknowledgments}
The authors would like to thank N. K. Sato and K. Deguchi for valuable discussions. 
N. T. is supported by JSPS KAKENHI Grant No. JP16H07447, JP19H05817, and JP19H05820.
S. S. is supported by JSPS KAKENHI Grant No. JP26800179, No. JP16H06345 and JP20H05279.
R.A. is supported by MEXT KAKENHI Grant No. JP15H05883 and JSPS KAKENHI Grant No. JP16H06345.
\end{acknowledgments}


\bibliographystyle{apsrev4-1}
\input{specific_heat.bbl}
\end{document}

%% file: specific_heat.bbl
%